\documentstyle[prb,aps,]{revtex}
\def\Dct{$\Delta_{ct}$}
\newcommand{\BSCCO}{Bi$_2$Sr$_2$Ca$_2$Cu$_2$O$_{8+\delta}$}
\newcommand{\LSCO}{La$_{2-x}$Sr$_x$CuO$_4$}
\newcommand{\LBCO}{La$_{2-x}$Ba$_x$CuO$_4$}
\newcommand{\NCCO}{Nd$_{2-x}$Ce$_x$CuO$_4$}
\newcommand{\YBCO}{YBa$_2$Cu$_3$O$_7$}
\newcommand{\BKBO}{Ba$_{1-x}$K$_x$BiO$_3$}

\def\@pnumwidth{2em}

\begin{document}

%\draft

\twocolumn[\hsize\textwidth\columnwidth\hsize\csname @twocolumnfalse\endcsname

\title{Insulator-metal transition in
high-T$_c$ superconductors as result of percolation over -U~centers}

\author{K. V. Mitsen and O. M. Ivanenko}
\address{Department of Solid State Physics,
P. N. Lebedev Physical Institute RAS, Moscow 117924, Russia}

\date{\today}

\maketitle

%\vspace{-0.1in}

%%%%%%% abstract (below) %%%%%%%%%%%%%
\begin{abstract}
The mechanism of -U center formation in high-$T_{c}$ superconductors (HTS) with doping is considered. It is shown
that the transition of HTS from insulator to metal passes through the particular dopant concentration range where
the local transfer of singlet electron pairs from oxygen ions to pairs of neighboring cations (-U centers) are
allowed while the single-electron transitions are still forbidden. We believe it is this concentration range that
corresponds to the region of high-Tc superconductivity and the interelectron attraction results from the
interaction of electron pairs with -U centers. Additional hole carriers are generated as the result of singlet
electron pair transitions from oxygen ions to -U-centers. The orbitals of the arising singlet hole pairs are
localized in the nearest vicinity of -U center. In such a system the hole conductivity starts up at the dopant
concentration exceeding the classical 2D-percolation threshold for singlet hole pair orbitals. In the framework of
the proposed model the phase diagram Ln-214 HTS compounds is constructed. The remarkable accord between
calculated and experimental phase diagrams may be considered as the confirmation of the supposed model. The main
features of hole carriers in HTS normal state are found to be the nondegenerate distribution and the dominant
contribution of electron-electron scattering to the hole carrier relaxation processes. Various experimentally
observed anomalies of HTS properties are shown to be the consequences of the above-mentioned features. The
conclusion is made that HTS compounds are the special class of solids where the unusual mechanism of
superconductivity different from BCS is realized.
\end{abstract}
 \vskip1.5pc]
%%%%%%% abstract (above) %%%%%%%%%%%%%

\section{Introduction}
\label{sec:intro}

%\begin{twocolumn}

%%%%%%%%%%%%%%%%%%%%%%% main text (below) %%%%%%%%%%%%%%%%%%%

In 13 years elapsed after the discovery of HTS,\cite{Bednorz} numerous models have been suggested (see review in
Ref.\ \onlinecite{Loktev}) to explain the nature of the ground state and anomalous properties of HTS compounds.
However, the lack of any crucial experiment gave no way of choosing between these models.

In this paper, we intend to demonstrate that the mechanism responsible for various anomalous properties of HTS
compounds (including the high-temperature superconductivity itself) is apparently based on interaction of
electrons with so-called -U centers.\cite{Anderson} To this end, we consider the way the insulator-metal
transition occurs in HTS under doping. On the basis of a simple ionic model, we will show that such a transition
should pass through a certain range of dopant concentrations; this range corresponds to the situation when local
transitions of singlet electron pairs from oxygen ions to a pair of neighboring cations (-U center) become
possible in individual microclusters including several unit cells, while the single-electron transitions are
still forbidden. In our opinion, it is this range of concentrations that corresponds to the HTS range where
interelectron attraction results from interaction of electron pairs with -U
centers.\cite{Simanek,Ting,Schuttler,Massida,Volkov,Eliashberg,Kulik} Conduction in such a system arises when the
concentration of -U centers exceeds the percolation threshold for orbitals of singlet hole pairs. We are going to
consider which specific fragments of crystal structure are involved in the formation of -U centers and what is
the range of existence of infinite percolation cluster interconnecting singlet hole pair orbitals. On this basis
we will construct the phase diagram for Ln-214 (Ln=La, Nd) compounds. In our opinion, a comparison of this phase
diagram with the carefully investigated diagram for Ln-214 compound should be taken as the aforementioned crucial
experiment for choosing the mechanism responsible for the HTS properties. We discuss the special features of
phase diagrams for Ln-214 with $n$- and $p$-type of doping, and also we give a somewhat different interpretation
of some experimental results obtained for underdoped HTS. Furthermore, we consider the nature of mobile hole
carriers and mechanism of their relaxation in HTS. As will be seen the distinctive feature of HTS normal state is
the nondegenerate distribution of hole carriers that results in hole-hole scattering to dominate in kinetic
processes. It will be shown that this feature may account for the unusual transport and optical properties of HTS
observed in experiment.

\section{-U center formation in HTS}

\label{sec:spectr}

There are good reasons to believe that electronic spectrum of insulator phase for various HTS compounds in the
vicinity of Fermi energy E$_F$ can be best approximated by the model of charge-transfer insulator.\cite{Zaanen}
In this model, the upper empty band formed by unfilled orbitals of cations is separated with a gap from the O2p
band formed largely by oxygen states (Fig.\ \ref{fig1}(a)). The gap {$\Delta_{ct}$} existing in the spectrum is
related to the transfer of electron from oxygen to the neighboring cation and lies in the range of $1.5-2$ eV for
all HTS.\cite{Ohta}

The question arises: What is the mechanism of insulator-metal transition in the doped HTS? As an example of HTS,
we consider Ln-214. For these compounds, the quantity $\Delta _{ct}$, in terms of simple ionic model, is defined
by the following relationship:\cite{Mazumdar}

\begin{equation}
\Delta _{ct}\approx \left| \Delta E_M\right| +A_p-I_d.  \label{ct}
\end{equation}

Here, $I_d$ is the second ionization potential of Cu, $A_p$ is the
electronegativity of oxygen with respect to formation of O$^{2-}$, and $%
\left| \Delta E_M\right| $ is the difference of the Madelung energy $E_M$ between the configuration in which the
copper and oxygen atoms are in the
state of Cu$^{2+}$ and O$^{2-}$ and that with these atoms in the states of Cu%
$^{1+}$ and O$^{1-}$. Taking into account\cite{Eschrig} that $I_d\sim 20$ eV, $A_p\sim 0$ eV and $\Delta_{ct}\sim
1.5-2$ eV, there is a subtle balance between these three quantities.

This balance can be varied by heterovalent doping; for example by doping the La$_2$CuO$_4$ with divalent Sr or by
doping Nd$_2$CuO$_4$ with tetravalent Ce. It is a matter of great importance that the charge carriers introduced
by doping (so-called doped carriers) are localized \cite{Romberg,Martin,Hammel} (at least for low concentrations)
in CuO$_2$ plane in the vicinity of the dopant ion: either at the O2p$_{x,y}$ orbitals (the holes in
La$_{2-x}$Sr$_x$CuO$_4$) or at the Cu3d$_{x^2-y^2}$ orbitals (electrons in Nd$_{2-x}$Ce$_x$CuO$_4$). Taking into
account that the interaction of O$^{2-}$ and Cu$^{2+}$ gives the main contribution to $E_M$, an addition both of
electrons (to the Cu orbitals) and of holes (to the oxygen orbitals) will results in the same thing: a decrease
in $\left| \Delta E_M\right| $ and, correspondingly, a decrease in {$\Delta _{ct}$} for other pairs of copper and
oxygen ions sited in the vicinity of the doped carrier. For a certain critical concentration $x_c$, the gap
{$\Delta _{ct}$} vanishes throughout the entire crystal. As a result, electron transitions from oxygen to copper
become possible, and the material passes into a normal metal.

In generally by this means, we can conceive the transition of a charge-transfer insulator into the metallic state
with doping in terms of ionic model. However, we argue that, in HTS compounds, the transition from insulator to
metal with x passes through a special range of concentrations $x_0<x<x_c$ for which the two-electron transitions
from oxygen ions to certain pairs of neighboring cations become possible, while the single-electron transitions
are still forbidden. In other words, -U centers are formed on certain pairs of cations at $x_0<x<x_c$.

Let us consider a Cu$_2$M$_2$O$_n$ cluster where two neighboring Cu ions belong to the CuO$_2$ plane, and M=Cu in
the selfsame CuO$_2$ plane for Ln-214, M=Cu
in the CuO$_3$ chain for YBa$_2$Cu$_3$O$_7$, and M=Bi for Bi$_2$Sr$_2$Ca$_2$%
Cu$_2$O$_{8+\delta }$. As will be seen from the following consideration, the condition for the formation of -U
centers at the neighboring Cu ions in CuO$_2$ plane consists in the presence of one doped carrier in the vicinity
of each M ion (in {\YBCO} and {\BSCCO}, these carriers are in the CuO$_3$ chains and BiO planes, respectively).
In Ln-214, two types of such clusters are possible (Figures\ \ref{fig2}(a, b)): the projections of the dopant
ions onto the CuO$_2$ plane are separated by either 3$a$ or $a\sqrt{5}$, where $a$ is the lattice constant in the
CuO$_2$ plane. In both cases, the presence of doped carrier in the vicinity of each M ion reduces {$\Delta
_{ct}$} for the neighboring Cu ions and, as it will be shown, provides the conditions (i.e., forms a local
minimum of potential energy) for simultaneous transition of two electrons to the pair of interior Cu ions from O
ions surrounding this pair. It should be noted here that, in the intermediate case where the M ion projections
are spaced $a\sqrt{8}$ apart, such pair of Cu ions does not emerge (Fig.\ \ref{fig2}(c)).

It is possible to estimate a decrease in {\Dct} for a given Cu ion in La-214 due to the presence of a single hole
around the neighboring Cu ion if we assume that this hole is ''distributed'' (Fig.\ \ref{fig3}) over 12
nearest-neighbor oxygen ions (the first and second coordination spheres). This assumption is consistent with
experimentally determined limit of substitutability (see below). We take into account only the interaction
between the nearest neighbors. Therefore we consider 3 oxygen ions from the total number of 12, where the hole
''feels'' the unscreened Cu ion separated by a distance of $r=a/2\approx 2$ {\AA } from this hole. In this case
for a given Cu ion, a lowering of the Cu3d$^{10}$ state energy amounts to $\Delta {E}\sim (1/4)e^2/r\sim 1.8$ eV;
here, $e$ is the electron charge. This means that, due to doping, the Cu3d$^{10}$ state energy for a given Cu ion
is decreased, so that it is now by $\Delta {E}\sim 1.8$ eV below the bottom of the conduction band for undoped
material. This value is still smaller by $\sim 0.1-0.2$ eV than $\Delta _{ct}\approx  1.9-2.0$ eV for
La$_2$CuO$_4$.\cite{Ohta} In Nd-214 doped electrons increase O2p state energy around M ion that results in a
decrease in {\Dct}, too.

An additional lowering of the Cu3d$^{10}$ state energy is attained owing to formation of a bound state of two
electrons at neighboring Cu ions in the presence of two holes in the immediate vicinity of this pair. Such a
decrease is possible for a bonding orbital of a singlet hole pair, as is the case for H$_2$ molecule. Here, this
analogy is even more appropriate because
the distance between electrons at Cu ions ($\approx 3.8$ \AA ) is close to $%
R_0\epsilon _\infty \approx 3.6$ \AA , where $\epsilon _\infty \approx 4.5$ is the high-frequency dielectric
constant\cite{Harshman} and $R_0\approx 0.8$ {\AA} is the distance between nuclei in H$_2$ molecule. Therefore, an
additional lowering of the energy $\delta E_U$ due to transition of two electrons to neighboring copper ions
can be, in the case under consideration, estimated from the relationship $\delta {E}_U\sim \delta {E}%
_H/{\epsilon _\infty ^2}\approx $ 0.23 eV, where $\delta {E}_H=4.75$ eV is the binding energy in H$_2$ molecule.
However, this value $\delta {E}_U$ is apparently underestimated because oxygen ions in CuO$_2$ plane efficiently
screen the repulsive interaction of electrons at Cu ions and weakly screen the electron-hole attraction.

Thus, we may assume that $\Delta _{ct}$, which amounts to $1.5-2.0$ eV for doped cuprates,\cite{Ohta} vanishes for
two-electron transitions to neighboring Cu ions. In this case, holes apparently occupy predominantly $\pi
{p_{x,y}}$ orbitals, thus providing naturally the bonding character of the hole-pair orbital owing to
configuration of bonds in the CuO$_2$ plane, which allows the holes to reside in the space between Cu ions (Fig.\
\ref{fig4}). Here we assume that the states at the top of the oxygen valence band are formed predominantly by
$\pi p_{x,y}$ orbitals.\cite{Guo}

It follows from the above consideration that, if an appropriate local minimum in {$\Delta _{ct}$} is formed, the
bound electron state can arise at the pair of neighboring Cu ions while the single-electron transitions are
forbidden (i.e., -U center is formed).\cite{BKBO1} In this case, a singlet hole pair is localized in the vicinity
of  -U center at a distance of $\sim {a/2}$. The region of the hole-pair localization is restricted by the
condition that the pair level lies in line with the top of the valence band (the energy of the pair level becomes
higher with increasing of the hole-pair localization area). Thus, the energy spectrum of CT-insulator under
doping is modified by the addition of electron pair level lying in line with the top of the valence band (Fig.\
\ref{fig1}(b)). In this case the states at the top of valence band should be consider as formed by the $\pi
p_{x,y}$ orbitals of oxygen ions surrounding -U centers.

Conduction in such a system occurs if the areas of the  localization of singlet hole pairs form a percolation
cluster. The localization areas of doped holes may enter into this percolation cluster, too (e.g. in La-214). The
percolation threshold for the hole-pair orbitals of the -U centers in Ln-214 is coincident with that for a
ensemble of segments of length $L=3a$ or $a\sqrt{5}$ in square lattice. The region of delocalization of hole
carriers in the percolation cluster is also restricted by the condition that the position of the pair level
should coincide with the valence-band top. Such a mechanism keeps the pair level exactly at the top of the
valence band (Fig.\ \ref{fig1}(b)).

On the other hand, if projections of two M ions are separated by a distance 2$a$ (Fig.\ \ref{fig2}(d)), {$\Delta
_{ct}$} vanishes for single-electron transitions to interior Cu ion as well.\cite{BKBO2} Such a fragment is a
nucleus of ordinary metallic phase. For the relevant concentrations $x\geq x_c$, the entire crystal transforms
into ordinary metal. This state corresponds to a single-band electron spectrum. In {La$_{2-x}$Sr$_x$CuO$_4$} in
the ordinary metal phase, the charge carriers are electrons because, due to doping with divalent Sr, the filling
of the band $\rho <1/2$; under the same conditions, the charge carriers are holes in {Nd$_{2-x}$Ce$_x$CuO$_4$}
because, due to doping with tetravalent Ce, we have $\rho >1/2$. The intermediate range of $x$ where -U centers
and ordinary metal phase are coexisted in the percolation cluster is so-called ''overdoped'' region. In this case,
the additional carriers from metal phase decrease $\left| \Delta E_M\right| $ and lower the pair level below the
top of the valence band.

\section{Construction of phase diagram of L\lowercase{n}-214 compounds}

\label{sec:fd}

We will construct the phase diagram for Ln-214 system using the following assumptions:

(i) the -U centers are formed at the pairs of neighboring Cu ions belonging only to the clusters with $L=3a$ and
$L=a\sqrt{5}$ ;

(ii) the orbitals of hole pairs are in the immediate vicinity of these pairs at a distance of $a/2$;

(iii) conduction in the system arises in the case of percolation over segments containing -U centers; and

(iv) localized doping carriers cannot be separated by a distance smaller than 2a.

Assumption (iv) follows from the existence of the substitutability limits in Ln-214 $x_{lim}\approx
0.2-0.25$.\cite{Radaelli,Yoshimura,Paulus} If these limits are exceeded, a decomposition of single-phase state
and/or a change of the oxygen content occur. In our opinion, the existence of the substitutability limit is
related to repulsive interaction between localized doped carriers. In turn, such a repulsion would affect the
dopant ion distribution, if the mobility of these ions at heat treatment temperature was rather high. Therefore,
we believe that the dopant ions (more precisely, their projections onto the CuO$_2$ plane) cannot be separated by
a distance $L<2a$ as well.

Taking into account the above assumption, we may define the $2D$-percolation threshold as follows. Let us assume
that we have a square lattice with the cell parameter $a$=1 and let a fraction $x$ of the sites be occupied by
atoms. In order to determine the percolation threshold for segments with length $L$ (i.e., for the pairs of atoms
separated by a distance $L$) we locate each occupied site at the center of circle with the radius $L/2$ (Fig.\
\ref{fig5}(a)). The sum of the areas of the circles constructed around these atoms is given by $S=x{\pi }L^2/4$.
For a square lattice, the percolation sets in when $S\geq $ 0.466.\cite{Ziman} Consequently, the concentration
corresponding to the percolation threshold is given by $x_p(L)=0.593/L^2$. Here, we assume that the distribution
of atoms over the sites is random, and L is the smallest distance between atoms for the given concentration.
Otherwise, we would have an overlapping of circles (Fig.\ \ref{fig5}(b)), and the value of $x_p$ would be larger
than that obtained from the above relationship. In this case a percolation cluster would include not only the
segments with a length of L but also those with smaller length (linking the sites separated by smaller
distances). The maximal number of segments with length $L$ - $x_M(L)$ - can be attained in the case of ordered
arrangement of atoms in a square lattice with the parameter $L$; i.e., $%
x_M(L)=1/L^2.$ The values of $x_p$ and $x_M$ for various values of $L$ are listed in Table 1. Therein, in the
righthand column, it is indicated which state (insulator, metal, or HTS) would correspond to the case of
percolation over the segments with a given $L$.

Figure\ \ref{fig6}(a) shows the percolation ranges for segments with different length $L$ (therein, the
corresponding value of $L^2$ is indicated to left of each rectangle). The left side of each rectangle corresponds
to the $2D$-percolation threshold for the segments with the length $L$ under the condition that the atoms are
randomly distributed and there are no segments with the length smaller than $L$. The right side of each rectangle
corresponds to the point $x_M$. Heavy lines indicate the percolation ranges for the segments with $L=3$ and
$L=\sqrt{5}$. It is in these concentration regions we suggest, the high-temperature superconductivity occurs.

As will be evident from what follows, the ratio between order and disorder in the distribution of dopants over
the sites is very important and defines all special features of phase diagrams for HTS of Ln-214 type. We believe
that a tendency towards ordering is related to the difference in ionic radii of Ln and the dopant. This tendency
should be most pronounced for the La/Ba pair and least pronounced for the Nd/Ce pair. The degree of ordering
should also increase with increasing $x$ (with decreasing $L$). In what follows, while on subject of the dopant
ordering, we imply a formation of a large number of ordered clusters (with sizes smaller than 100 \AA ) separated
by thin walls with random distribution of the dopant. In these walls, the constraint $L\geq 2$ is also preserved.
Thus, only the short-range order exists in the system. It is assumed that, if the value of $x_M$ is exceeded, a
transition to random distribution of atoms over the sites occurs with formation of an infinite percolation
cluster with a smaller value of $L$.

We now consider various ranges of concentrations in Fig.\ \ref{fig6}(a) assuming that, at each point $x>0.1$, no
more than two types of segments exist.

1. $0.20<x<0.25$. In this case, the $2D$-percolation is effected over the segments with $L=2$, i.e., over the
clusters of ordinary metal.

2. $0.148<x<0.20$. In this range, the $2D$-percolation threshold for the segments with $L=2$ depends on the
degree of ordering of the dopant atoms with $L=\sqrt{5}$ . In the case of ordering of the atoms with $L=\sqrt{5}$,
percolation over the segments with $L=2$ sets in at $x=0.2$, whereas, for random distribution, percolation over
the segments with $L=2$ is attained at $x=0.148$. Therefore, this range of $x$ corresponds either to HTS, or to a
mixed state of HTS and ordinary metal.

3. $0.125<x<0.148$. Here, we have a "pure" $2D$-percolation over the segments with $L=\sqrt{5}$ . This range
corresponds to HTS.

4. $0.118<x<0.125$. If the atoms with $L=\sqrt{8}$ are ordered, percolation over the segments with $L=\sqrt{5}$
sets in at $x=0.125$, whereas, if the atoms with $L=\sqrt{8}$ and $L=\sqrt{5}$ are randomly distributed,
percolation over the segments with $L=\sqrt{5}$ sets in at $x=0.118$. In this range, we have either insulator (in
the former case) or HTS (in the latter case).

5. $0.111<x<0.118$. This is the domain of "pure" $2D$-percolation over the segments with $L=\sqrt{8}$. This phase
corresponds to an insulator.

6. $0.10<x<0.111$. In the case of ordering of atoms with $L=3$, the $2D$%
-percolation over the segments with $L=3$ sets in at $x=0.10$, whereas, if the atom pairs with $L=3$ and
$L=\sqrt{8}$ are randomly distributed, there is no $2D$- percolation neither over the two types of segments. In
this case, a percolation cluster would include the areas with L=3 (HTS) and with $%
L=\sqrt{8}$ (insulator) and the conduction is possible only through the tunneling between clusters with $L=3$
(3D-percolation).

7. $0.066<x<0.10$. In this range, there is no $2D$-percolation over the segments with $L=3$. $3D$-percolation
(and superconductivity) is still possible for $x>0.077$, whereas, for $x<0.077$, there is no bulk
superconductivity.

For comparison, Figures\ \ref{fig6}(b - d) show the experimental phase diagrams $T_c(x)$ for
{La$_{2-x}$Ba$_x$CuO$_4$},\cite{Moodenbaugh} {La$_{2-x}$Sr$_x$CuO$_4$},\cite{Kumagai} and Nd$_{2-x}$Ce$_x$CuO$_4$.
\cite{Takagi} Comparing Figures\ \ref{fig6}(a - d), we can readily see that all particular points in experimental
phase diagrams practically coincide with the boundaries of percolation domains corresponding to the segments with
different values of $L$.
Difference between the phase diagrams of {La$_{2-x}$Ba$_x$CuO$_4$} and {La$_{2-x}$Sr$_x$%
CuO$_4$} consists in the fact that, in the former case, a dip in $T_c$
occurs at $x_d=0.125$, whereas, in the latter case, it takes place at $%
x_d=0.115$. It is reasonable to relate the difference in the values of $x_d$ for these compounds to a larger
degree of ordering in the La/Ba sublattice as compared to that for La/Sr, as a result of which the percolation
threshold for the segments with $L=\sqrt{5}$ becomes shifted to $x=0.125$ (the point of the highest ordering for
$L=\sqrt{8}$).

The maximum of $T_c$ at $x\approx 0.15$ is related to the fact that the ordinary metal clusters appear within the
superconducting phase for $x>0.148$. Because of this, superconductivity is ''depressed'' to a degree
corresponding to the ratio of the volumes occupied by superconducting and metal (nonsuperconducting) phases. This
is confirmed \cite{Kitazawa} by measurements of the Meissner-phase volume as a function of a magnetic field.
These measurements show that, for $x=0.15$, this volume is virtually independent of the field, whereas, for
$x>0.15$, a magnetic field reduces significantly the volume of superconducting phase. At the same time, in low
fields, the magnitude of the Meissner effect increases \cite{Radaelli,Kitazawa} with $x$ up to $%
x\approx 0.2$, which is indicative of ordering of the dopants with $L=\sqrt{5}$.

In {\NCCO} (Fig.\ \ref{fig6}(d)), there is virtually no ordering of Ce owing to a small differences in sizes of Nd
and Ce ions. Therefore, the $2D$-percolation is possible only for $x>0.118$ (for $L=\sqrt{5}$).\cite{NCCO}
However, due to the absence of ordering, a percolation
cluster would also include the regions with $L=2$; thus, the percolation threshold for $L=\sqrt{%
5}$ shifts to that for $L=2$. This is consistent with experimental phase diagram.

The region $x<0.12$ for {La$_{2-x}$Ba$_x$CuO$_4$} and {La$_{2-x}$Sr$%
_x$CuO$_4$} deserves special consideration (the region of ''underdoping'').
As it follows from Fig.\ \ref{fig6}a, even the $3D$-percolation is not possible for $%
x<0.077$, and only the ''traces'' of superconductivity can be observed. This conclusion is consistent with the
results reported in \cite{Yoshizaki,Marin} where the bulk superconductivity was not observed in {La$_{2-x}$Sr$_x
$CuO$_4$} for $x=0.08$. As mentioned above, it is not to be expected that the $2D$-percolation can originate (at
least, in La$_{2-x}$Sr$_x$CuO$_4$) in the range of $0.08<x<0.12$ as well because the percolation regions for
three different types of segments with $L^2=8,9,$ and $10$ are close to each other. Most likely, we can have here
a tunneling between clusters with $L=3$. This inference accounts for the results reported in Ref.\
\onlinecite{Ando} where, in {\LSCO} at $T\rightarrow 0$, a logarithmic divergence of resistivity was observed for
$x<0.15$, with superconductivity being depressed by a magnetic field.

The proposed model allows to give an alternate interpretation of the experiments on the pseudogap observation in
under- and optimally- doped HTS.\cite{Ding,Marshall,Ding2} As it follows from experiment the pseudogap has the
same symmetry and about the same value as the superconducting gap but it vanishes at $T^{*}>T_c$ ($T^{*}$ decrease
with $x$ down to $T_c$).

We believe that the observed pseudogap is nothing but just the same superconducting gap opening at $T>T_c$
because of the large fluctuations of the number of particles between valence band and pair level in the short
chains of -U centers. The point is that in HTS the mechanism of the superconducting gap suppression is the
occupation of pair level with electrons. Therefore the decrease of pair level occupation by fluctuation may
result in turning on superconductivity in such chains at $T^{*}>T_c$ (first-order phase
transition.\cite{Ivanenko}) At small $x$ the great bulk of -U centers are grouped on short chains where large
relative fluctuations of the number of particles are possible. With $x$ the increasing part of -U centers belong
to the infinite percolation cluster. Therefore $T^{*}$ decrease with $x$ down to $T_c$, unless all -U centers are
integrated in infinite percjlation cluster.

Thus, we may conclude that all special features observed in phase diagrams of Ln-214 HTS represent no more than
geometric relations in a square lattice and a competition between ordering and disorder in distribution of the
dopant ions. These items determine the range of existence of infinite percolation cluster integrating the
localized orbitals of singlet hole pairs of -U centers. The agreement between experimental and calculated phase
diagrams supports the conclusion that it is the considered fragments involving the pairs of neighboring Cu ions
in the CuO$_{2}$ plane which are responsible for superconductivity in Ln-214. It is interesting that the
percolation cluster in Ln-214 resembles the Little's «polymer» \cite{Little} while that in {\YBCO} and {\BSCCO}
(where doped holes are in the plane parallel to the CuO$_2$ plane) resembles Ginsburg's «sandwich».\cite{Ginsburg}

\section{Hole carrier generation and relaxation processes in HTS}

\label{sec:relax}

Now we consider the process of the generation of hole carriers as well as the features of the transport and
optical properties of HTS in the framework of the proposed model of electronic spectrum (Fig.\ \ref{fig1}(b)). Let
there is an infinite cluster including a quantity of -U centers together with the nearest oxygen ions. The
two-particle hybridization of the pair states with the O2$p_{x,y}$ states of oxygen ions surrounding -U centers
results in the broadening both of the pair states and of the O2$p_{x,y}$ ones. (In this case the electrons at the
top of the valence band resemble the ''marginal'' Fermi liquid).\cite{Varma} The pair level broadening can be
expressed \cite{Eliashberg,Kulik} as $\Gamma \sim \pi (DV)^2kT$, where $V$ is the constant of hybridization, D is
the density of states at the top of O2p band. The broadening of band states is $\gamma \sim \Gamma /DE_0$, where
$E_0 $ is the energy width of statistic distribution of pair states over sites. This broadening smoothes the
features of the band state density and results in its independence on energy over the interaction range. The
occupation of pair level results from transitions of electrons from $\pi p_{x,y}$ oxygen orbitals to the -U
centers and accompanied by the generation of hole carriers in the oxygen band. We call the phase where the
additional hole carriers appear through this mechanism as -U phase. The electron occupancy of -U centers $\eta$
as well as the hole concentration $n$ in -U phase are determined by the balance of rates of electron pair
transitions from oxygen band to pair level and back. If $N$ is -U center concentration so $n=2N\eta $. The rate
of electron transitions from pair level to singlet orbitals is proportional to $N\eta \Gamma \sim \eta T$. The
rate of reverse process is determined by the frequency of electron-electron scattering and proportional to $\gamma
^2(1-\eta )\propto T^2(1-\eta )$. Therefore
\begin{equation}
n=2NT/(T_0+T),  \label{n}
\end{equation}
where $T_0$ is temperature-independent value. Thus $n\propto T$ at low temperature and tends to $2N$ at high
temperature. This is in agreement with the data of Hall measurements in \YBCO, where doped carriers in chains do
not contribute to Hall conductivity.\cite{Kapitulnik,Ong} As it follows from the above consideration the hole
carrier distribution turns out to be nondegenerated owing to the interaction with -U centers. Taking into account
the nondegeneracy of carrier distribution (absence of the Pauly blocking) and their high concentration
$(10^{21}-10^{22}$\ cm$^{-3})$ the electron-electron scattering (more precisely to say hole-hole scattering in
this case) is likely to provide the dominant contribution to the relaxation processes in HTS. As far as the
interaction of two holes in the system with -U centers corresponds to the effective attraction, it will not be
conventional Coulomb scattering. The main mechanism of carrier relaxation in HTS is likely to be similar to that
assumed for the metals and alloys with strong electron-phonon coupling.\cite{MacDonald} In these materials the
electrons in a layer $\sim k\Theta _D$ at the Fermi surface ($\Theta _D$ is Debye temperature) are attracted due
to virtual phonon exchange. This electron-electron interaction enhanced by phonon-mediation effects far exceeds
the screened Coulomb repulsion. Therefore in the metals with strong electron-phonon interaction the dominant
channel of the electron-electron scattering will be determined by the virtual phonon exchange. The contribution
of these processes becomes essential\cite{MacDonald} for $T<\Theta _D$ with the amplitude of the
electron-electron scattering being independent of scattering particle energy $E$ for $E\ll k\Theta _D$ and
dropping sharply at $E\sim k\Theta _D$. For $E>k\Theta _D$ only the Coulomb interaction gives the contribution to
the amplitude of electron-electron scattering. The electron-electron contribution to the electrical resistivity
$\rho $ ($\rho =AT^2$) exceeding the electron-phonon one, have been observed experimentally for Al at $T<4$ K and
for the A15 superconductors at $T<50$ K,\cite{Garland,Gurvitch} with the amplitude $A$ more than one order
exceeding the result of calculations based on the Coulomb mechanism of scattering.

The dominant relaxation process in HTS is the electron-electron scattering as well and the main channel of this
scattering is the interaction of hole pairs on -U centers. This process may be considered as an exchange by
virtual bosons (exitons) with energy $W$. As far as $W\sim 0.1-1$ eV (in contrast to the exchange by virtual
phonon with energy $E\sim k\Theta _D<0.03$ eV) the temperature range, where the scattering processes with
exchange of virtual bosons are dominant, expands to $T\sim 10^3$ K. The temperature dependence $\rho (T)$ in such
a model may be obtained from the Drude formula $\rho =m^{*}\nu /ne^2$ (here $m^{*}$ is the effective mass of
holes, $\nu $ - the rate of hole-hole scattering). For $W\gg E$ the amplitude of scattering doesn't depend on the
particle energy. Assuming that density of states is energy-independent $\nu$ will be determined only by the hole
concentration and statistical factor in scattering cross-section (i.e. the phase volume available for occupation
with scattered particles). The latter is proportional to $E_1+E_2$ (here $E_1$ and $E_2$ are the energies of
scattering holes measured from the top of valence band). Thus
\begin{equation}
\nu \propto n(E_1+E_2).  \label{nu}
\end{equation}
For dc-conductivity $E_1\sim E_2\sim \gamma \propto T$ and $\nu \propto nT$, so that $\rho (T)\propto T$. This
kind of dependence has been observed
experimentally for the optimally doped samples of YBa$_2$Cu$_3$O$_7$, La$%
_{2-x}$Sr$_x$CuO$_4$, {Bi$_2$Sr$_2$Ca$_2$Cu$_2$O$_{8+\delta }$} and others. The dependence $\rho (T)=\rho _0+bT$
observed frequently in experiments may be explained by the contribution of insulating tunnel barriers
interlayering the areas of the -U phase that results in temperature-independent term $r_0$. For the overdoped
samples where both -U phase and metal phase coexist the additional carriers from metal phase decrease $\left|
\Delta E_M\right| $ and lower the pair level $\delta E$ below the top of the valence band. In this case the
distribution of hole carriers degenerates and $n$ is temperature-independent for $\gamma \ll \delta E$. The
temperature-dependent contribution in resistivity is $\rho (T)\propto \gamma ^{2} \propto T^2$. For $\gamma
\gtrsim \delta E$ the transition to the linear dependence $\rho (T)$ takes place.

The dominant contribution of electron-electron scattering into the scattering process also has an effect upon
both of frequency and temperature dependencies of optical conductivity $\sigma_{opt}$:
\begin{equation}  \label{opt}
\sigma_{opt}= \frac{e^{2}n}{m^{\ast}}\frac{\nu_{opt}}{{\omega}^{2}+{\nu_{opt}%
}^{2}}
\end{equation}
here $\omega$ is the photon frequency, $\nu_{opt}$ - the rate of optical
relaxation. For the electron-electron scattering (at $n=10^{21}-10^{22}$ cm$%
^{-3}$) the collision frequency is $\sim 10^{14}-10^{15}$ s$^{-1}$. Thus $%
\nu_{opt}\gg \omega$ for IR region and the formula for optical conductivity becomes more simple:
\begin{equation}  \label{opt2}
\sigma_{opt}(\omega,T)=e^{2}n/m^{\ast}\nu_{opt}.
\end{equation}

For optical relaxation we have $E_1\sim \omega , E_2\sim \gamma \propto T$. From where $\sigma _{opt}\propto
\omega ^{-1}$ (for $\omega \gg \gamma )$ and $\sigma _{opt}\propto T^{-1}$ (for $\omega <\gamma $). These results
are in a good agreement with the data of various experiments\cite{Schlesinger,Tanner}.

\section{Conclusions}

The mechanism of -U center formation in high-$T_{c}$ superconductors (HTS) with doping is considered. It is shown
that the transition of HTS from insulator to metal passes through the particular dopant concentration range where
the local transfer of singlet electron pairs from oxygen ions to pairs of neighboring cations (-U centers) are
allowed while the single-electron transitions are still forbidden. The additional hole carriers are generated as
the result of singlet electron pair transitions from the oxygen ions to the -U-centers. The orbitals of the
arising singlet hole pairs are localized in the nearest vicinity of -U center. In such a system the hole
conductivity starts up at the dopant concentration exceeding the classical 2d-percolation threshold for singlet
hole pair orbitals of -U centers. The main features of hole carriers in HTS normal state are found to be the
nondegenerate distribution and the dominant contribution of electron-electron scattering to the hole carrier
relaxation processes. These features account for the unusual transport and optical properties of HTS.

In the framework of the proposed model the phase diagram Ln-214 HTS compounds is constructed. The remarkable
accord between calculated and experimental phase diagrams may be considered as the confirmation of the supposed
model. Thus the HTS's may be considered as the special class of solids in-between insulator and metal where the
unusual mechanism of superconductivity resulting from interaction of electron pair with -U centers is realized.

\label{sec:concl}

\section{Acknowledgments}

We are grateful for valuable discussions with L.\ V.\ Keldysh, B.\ A.\ Volkov, E.\ G.\ Maksimov and P.\ A.\
Arseev.
%%%%%%%%%%%%%%%%%%%% main text (above) %%%%%%%%%%%%%%%%%

%%%%%%%% references (below) %%%%%%%%%%%%%

%%%%$%%%%%%%%%% references (above) %%%%%%%%%%%%%

\begin{figure}
\caption[] {(a) Electron spectrum of charge-transfer insulator in the vicinity of $E_F$; (b) Modification of
electron spectrum of a charge-transfer insulator as a result of doping (1 - the pair level of -U center).}
\label{fig1}
\end{figure}

\begin{figure}
\caption[~] {Four significant variants of relative positions of the nearest dopant ion projections. Here, the
fragments of crystal structure are on the left, and the corresponding projections onto the CuO$_2$ plane are on
the right. Ions of Sr (Ba, Ce) can also reside on opposite sides of the CuO$_2$ plane. In \NCCO, there are no
apical oxygen ions. 1 - -U centers, 2 - localization areas for doped carriers, 3 - Cu ion, for that {\Dct}
vanishes for one-electron transitions. (a, b) Two types of M$_{2}$Cu$_{2}$O$_{n}$ clusters that form the -U
centers at the interior Cu ions in {CuO$_2$} plane (for Ln-214, M=Cu in the selfsame {CuO$_2$} plane); (a) $L=3a$,
(b) $L=a\sqrt{5}$. (c) $L=a\sqrt{8}$. This relative position of the nearest projections of dopant ions corresponds
to insulator. (d) $L=2a$. This is an M$_{2}$Cu$_{2}$O$_{n}$ cluster serving as a «nucleus» of normal phase. }
\label{fig2}
\end{figure}

\begin{figure}
\caption[] {The area of the doped hole localization in the CuO$_2$ plane (shaded) involves 12 oxygen ions. Here,
crosses are copper ions, open circles are oxygen ions, and solid squares are Cu ions for which {\Dct} is
depressed by 1.8 eV owing to the presence of a hole at three neighboring oxygen ions.} \label{fig3}
\end{figure}

\begin{figure}
\caption[~] {Formation of bonding orbital (shaded) of a singlet hole pair of -U center from $\pi p_{x,y}$ oxygen
orbitals.} \label{fig4}
\end{figure}

\begin{figure}
\caption[~] {Constructions illustrating the method for determining the percolation threshold for segments with a
length of $L=\sqrt{5}$  in a square lattice. (a) Correct determination of the percolation threshold for segments
with length $L=\sqrt{5}$  in the case where the distance between atoms is no less than $L$; (b) Determination of
the percolation threshold for $L=\sqrt{5}$  if there are segments with $L=2$ yields an underestimated value of
$x_{p}$ because of overlapping of the circles. The segments with $L=2$ are shown by triple lines. The areas
corresponding to superposition of the circles are shaded.} \label{fig5}
\end{figure}

\begin{figure}
\caption[~] {(a) Boundaries of the percolation regions for the segments with various values of $L$. The left side
of each rectangle corresponds to the 2D-percolation threshold for the segments with length $L$ (under the
condition of random distribution of the circles with radius of $L/2$ over the sites and of the absence of the
segments with the length smaller than $L$). The right side of a rectangle is related to the point of the largest
number of segments with the length $L$ and corresponds to the ordered arrangement of atoms in a square lattice
with cell size $L$. The corresponding value of $L^2$ is indicated to the left of each rectangle. An increase in
the height of rectangles represents qualitatively an increase in the number of segments with decreasing $L$. (b -
d) Phase diagrams $T_{c}(x)$ for Ln-214 HTSs. The compositions for which superconductivity was not observed down
to 4.2 K are shown by triangles. (b) {\LBCO};\cite{Moodenbaugh} (c) {\LSCO};\cite{Kumagai} (d)
{\NCCO}.\cite{Takagi}} \label{fig6}
\end{figure}

%\clearpage
\begin{table}[t]
\caption[~]{Lower and upper bounds of concentration for percolation over segments with various L. In the
rightmost column, it is indicated which state (insulator, ordinary metal, or HTS) would correspond to a given
value of $L$ in the case of percolation over the segments with a given $L$. } \label{tab:tb1}
\begin{center}
\renewcommand{\arraystretch}{0}%
\begin{tabular}{llll}
\strut L$^2$ & $x_p$ & $x_M$ & Comments \\
\hline
\rule{0pt}{3pt}&\\
\strut 16 & 0.0371 & 0.0625 & insulator \\
\strut 13 & 0.0456 & 0.0769 & insulator \\
\strut 10 & 0.0593 & 0.100 & insulator \\
\strut 9 & 0.0659 & 0.111 & HTS \\
\strut 8 & 0.0742 & 0.125 & insulator \\
\strut 5 & 0.118 & 0.200 & HTS \\
\strut 4 & 0.148 & 0.250 & metal\\
\end{tabular}
\end{center}
\end{table}

\noindent \vspace{0.1in}


\begin{references}

\bibitem{Bednorz}  J. G. Bednorz and K. A. M\"{u}ller, Z. Phys. B {\bf 64},
189 (1986).
\bibitem{Loktev}  V. M. Loktev, Fiz. Nizk. Temp. {\bf 22}, 3 (1996) [Low
Temp. Phys. {\bf 22}, 211 (1996)].
\bibitem{Anderson}  P. W. Anderson, Phys. Rev. Lett. {\bf 34}${\bf ,}$ 953
(1975).
\bibitem{Simanek}  E. Simanek, Solid State Commun. {\bf 32}, 731 (1979).
\bibitem{Ting}  C. S. Ting, D. N. Talwar, and K. L. Ngai, Phys. Rev. Lett. {\bf %
45}, 1213 (1980).
\bibitem{Schuttler}  H.-B. Schuttler, M. Jarrell, and D. J. Scalapino, Phys.
Rev. Lett. {\bf 58}, 1147 (1987).
\bibitem{Massida}  J. Yu, S. Massida, A. J. Freeman, and D. D. Koelling,
Phys. Lett. A {\bf 122}, 203 (1987).
\bibitem{Volkov}  B. A. Volkov, V. V. Tugushev, Pis'ma Zh. Teor. Fiz. {\bf %
46}, 193 (1987) [Sov. Phys. - JETP Lett., {\bf 46}, 245 (1987)].
\bibitem{Eliashberg}  G. M. Eliashberg, Pis'ma Zh. Teor. Fiz. {\bf 46}
(supplement) 94 (1987) [Sov Phys. - JETP\ Lett. {\bf 46} (supplement), S81 (1987)].
\bibitem{Kulik}  I. O. Kulik, Fiz. Nizk. Temp. {\bf 13} 879 (1987) [Low Temp.
Phys. {\bf 13} 505 (1987)].
\bibitem{Zaanen}  J. Zaanen, G. A. ZSawatzky, and J. W. Allen, Phys. Rev.
Lett. {\bf 55}, 418 (1985).
\bibitem{Ohta}  Y. Ohta, T. Tohyama, and S. Maekava, Phys. Rev. Lett. {\bf 66%
}, 1228 (1991).
\bibitem{Mazumdar}  S. Mazumdar, Solid State Commun. {\bf 69}, 527 (1989).
\bibitem{Eschrig}  H. Eschrig, Physica C {\bf 159}, 545 (1989).
\bibitem{Romberg}   H. Romberg, M. Alexander, N. Nucker, P. Adelmann, and J. Fink, Phys. Rev. B%
{\bf \ 42}, 8768 (1990).
\bibitem{Martin}  R. L. Martin, Phys. Rev. B{\bf \ 53}, 15501 (1996)
\bibitem{Hammel} P. C. Hammel, R. L. Martin, B. W.Statt, F. C. Chou, D. C. Johnston, and S.-W. Cheong, Phys Rev B{\bf %
\ 57}, 712 (1998).
\bibitem{Harshman}  D. R. Harshman and A. P. Mills, Jr., Phys. Rev. B {\bf 45}%
, 10684 (1992).
\bibitem{Guo}  Y. Guo, J. M. Langlois, and W. A. Goddard III, Science {\bf 239}, 896 (1988).
\bibitem{BKBO1} In \BKBO, the -U centers are formed at the neighboring Bi
cations if there are three K ions in the eight cells surrounding each of these cations; i.e., each hole lowers
{\Dct} $\approx 2$ eV by $\sim 0.6$ eV. The less pronounced effect of hole in {\BKBO} as compared to La-214 is
accounted for by the fact that the negatively charged (with respect to Ba) K ion is in close vicinity to Bi ion.
\bibitem{BKBO2} In \BKBO, this occurs if there
are four K ions in the eight cells that surround a Bi ion.
\bibitem{Radaelli} P. G. Radaelli, D. G. Hinks, A. W. Mitchell, B. A. Hunter, J. L. Wagner, B. Dabrowski,
K. G. Vandervoort, H. K. Viswanathan, J. D. Jorgensen, Phys. Rev. B {\bf 49}, 4163 (1994).
\bibitem{Yoshimura} K. Yoshimura, H. Kubota, H. Tanaka, Y. Date, M. Nakanishi, T. Ohmura, N. Saga,
T. Sawamura, T. Uemura, K. Kosuge, J. Phys. Soc. Jpn. {\bf 62} 1114 (1993).
\bibitem{Paulus} E. F. Paulus, I. Yehia, H. Fuess, J. Rodriguez, T. Vogt,
J. Strobel, M. Klauda, G. Saemann-Ischenko, Solid State Commun. {\bf 73} 791 (1990).
\bibitem{Ziman}  J. M. Ziman. Models of disorder. The theoretical physics of
homogeneously disordered systems.-L.-N.Y.-M.: Cambridge University Press, 1979.
\bibitem{Moodenbaugh}  A. R. Moodenbaugh, Y. Xu, M. Suenaga, T. J. Folkerts,
R. N. Shelton, Phys. Rev. B {\bf 38}, 4596 (1988).
\bibitem{Kumagai} K. Kumagai, K. Kawano, I. Watanabe, K. Nishiyama, K. Nagamine, J. Supercond.
{\bf 7}, 63 (1994).
\bibitem{Takagi}  H. Takagi, S. Uchida, and Y. Tokura, Phys. Rev. Lett. {\bf %
62}, 1197 (1989).
\bibitem{Kitazawa}  K. Kitazawa, Y. Tomioka, T. Nagano, and K. Kishio, J.
Supercond. {\bf 7}, 27 (1994).
\bibitem{NCCO} The failure of 2D-percolation over the segments with $L=3$ in
{\NCCO} is also connected with the absence of doped holes «conglutinating» the singlet hole orbitals of -U
centers.
\bibitem{Yoshizaki} R. Yoshizaki, N. Ishikawa, H. Sawada, E. Kita, A. Tasaki, Physica C
{\bf 166}, 417 (1990).
\bibitem{Marin}  C. Marin, T. Charvolin, D. Braithwaite and R. Calemczuk,
Physica C {\bf 320}, 197 (1999).
\bibitem{Ando} Y. Ando, G. S. Boebinger, A. Passner, T. Kimura, K. Kishio, Phys. Rev.
Lett. {\bf 75}, 4662 (1995).
\bibitem{Ding} H. Ding, T. Yokoya, J. C. Campuzano, T. Takahashi, M. Randeria, M. R. Norman,
T. Mochiku, K. Kadowaki, and J. Giapintzakis, Nature {\bf 382}, 51, (1996).
\bibitem{Marshall} S. Marshall, D. S. Dessau, A. G. Loeser, C-H. Park, A. Y. Matsuura,
J. N. Eckstein, I. Bozovic, P. Founier, A. Kapitulnik, W. E. Spicer, Z. -X. Shen, Phys. Rev. Lett. {\bf 76}, 4841
(1996).
\bibitem{Ding2} H. Ding, M. R. Norman, T. Yokoya, T. Takuechi, M. Randeria,
J. Campuzano, T. Takahashi, T. Mochiku, and K. Kadowaki, Phys. Rev. Lett. {\bf 78}, 2628 (1997).
\bibitem{Ivanenko}  O. M. Ivanenko and K. V. Mitsen, Physica C {\bf 235-240},
2361 (1994).
\bibitem{Little}  W. A. Little, Phys. Rev. {\bf 134}, A1416 (1964).
\bibitem{Ginsburg}  V. L. Ginsburg, Zh. Eksp. Teor. Fiz. {\bf 47}, 2318
(1964) [Sov. Phys. - JETP {\bf 20}, 1549 (1965)].
\bibitem{Varma} C. M. Varma, P. B. Littlewood, S. Schmitt-Rink, E. Abrahams and A. E. Ruckenstein,
Phys. Rev. Lett. {\bf 63}, 1996 (1989).
\bibitem{Kapitulnik}  A. Kapitulnik, Physica C {\bf 153-155}, 520 (1988).
\bibitem{Ong}  N. G. Ong, T. W. Jing, T. R. Chien, Z. Z. Wang, T. V. Ramakrishnan, J. M.Tarascon, and K. Remschnig,
Physica C {\bf 185-189}, 34 (1991).
\bibitem{MacDonald}  A. H. MacDonald, Phys. Rev. Lett. {\bf 44}, 489 (1980).
\bibitem{Garland}  J. C. Garland and R. Bowers, Phys. Rev. Lett. {\bf 21},
1007 (1968).
\bibitem{Gurvitch} M. Gurvitch, A. K. Ghosh, B. L. Gyorffy, H. Lutz, O. F. Kammerer, J. S. Rosner, and M. Strongin, Phys.
Rev. Lett. {\bf 41}, 1616 (1978).
\bibitem{Schlesinger} Z. Schlesinger, R. T. Collins, F. Holtzberg, C. Feild, G. Koren, A. Gupta,
Phys. Rev. B {\bf 41}, 11237 (1990).
\bibitem{Tanner} D. B.Tanner, M. A. Quijada, D. N.Basov, T. Timusk, R. J.Kelley, M. Onellion, B. Dabrowski,
J. P. Rice, D. M.Ginsberg, J. Supercond. {\bf 8}, 563 (1995).

\end{references}
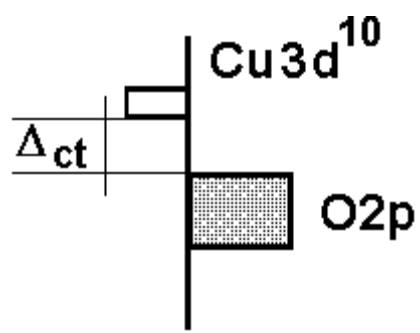
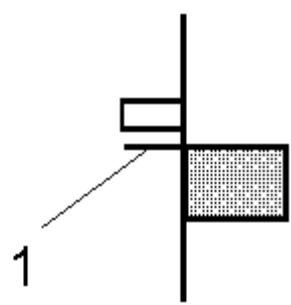
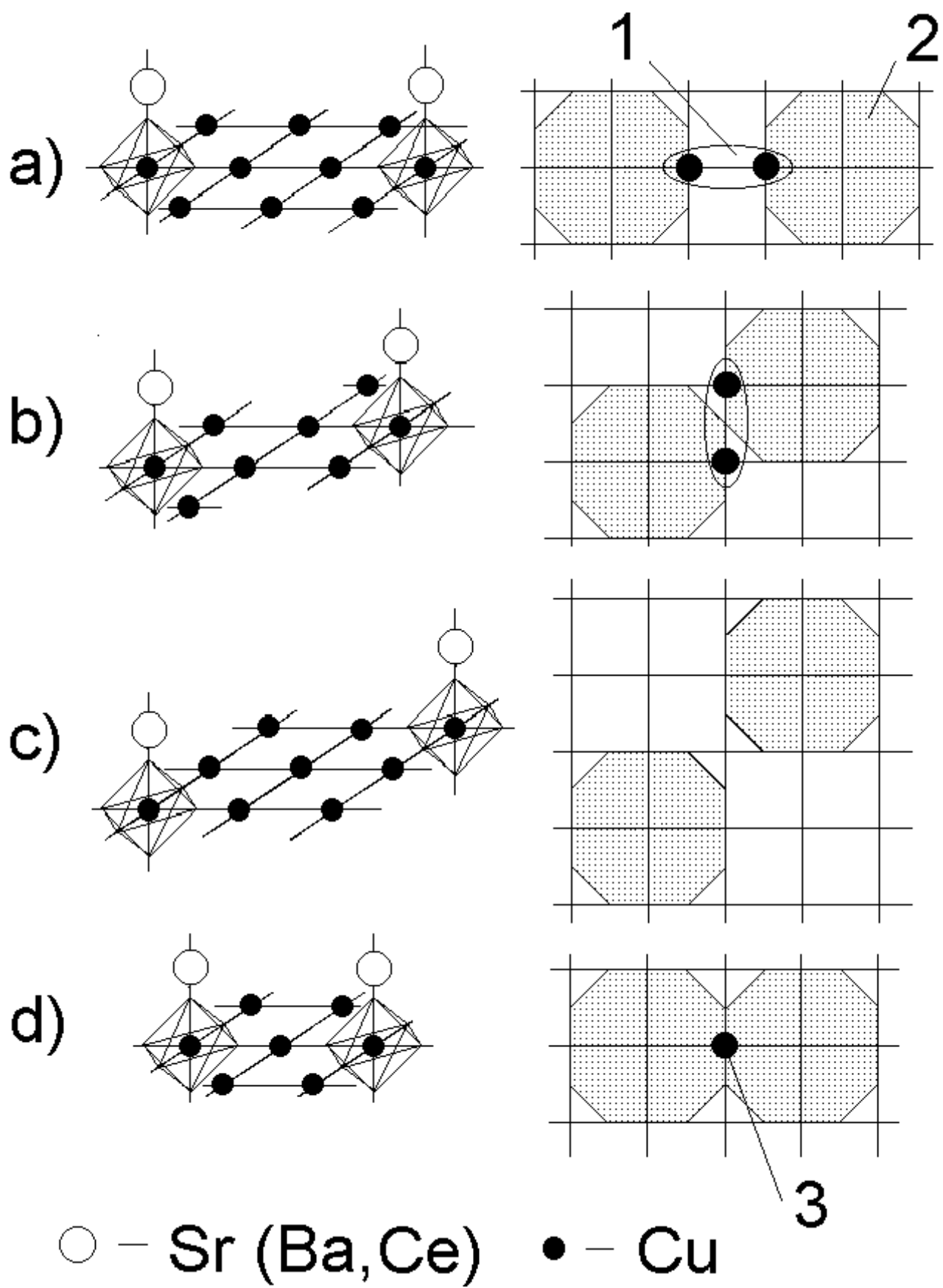
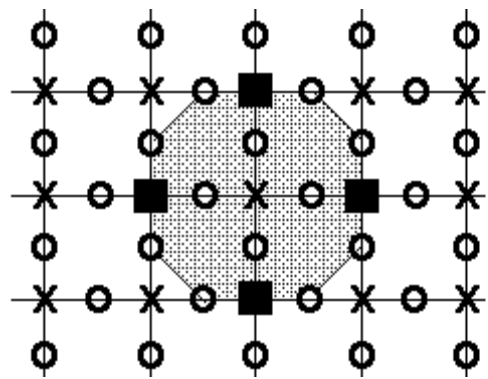
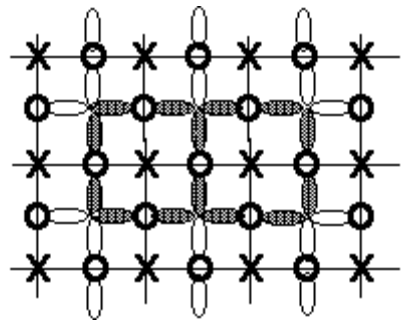
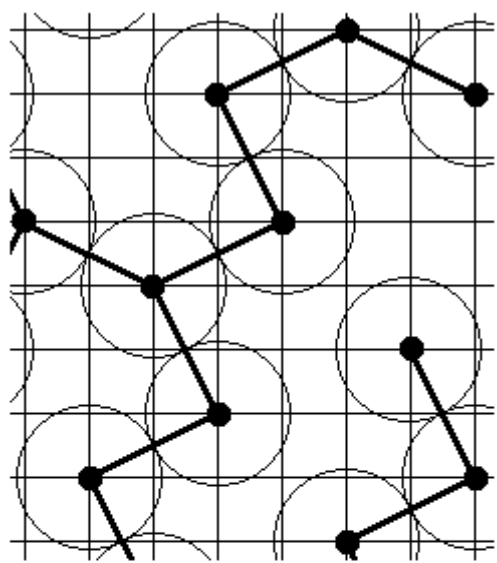
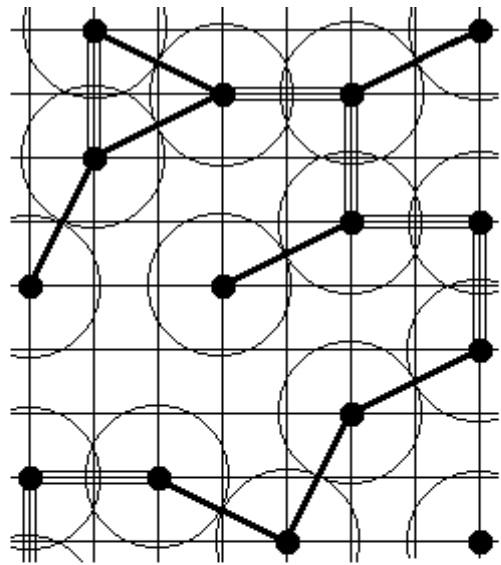
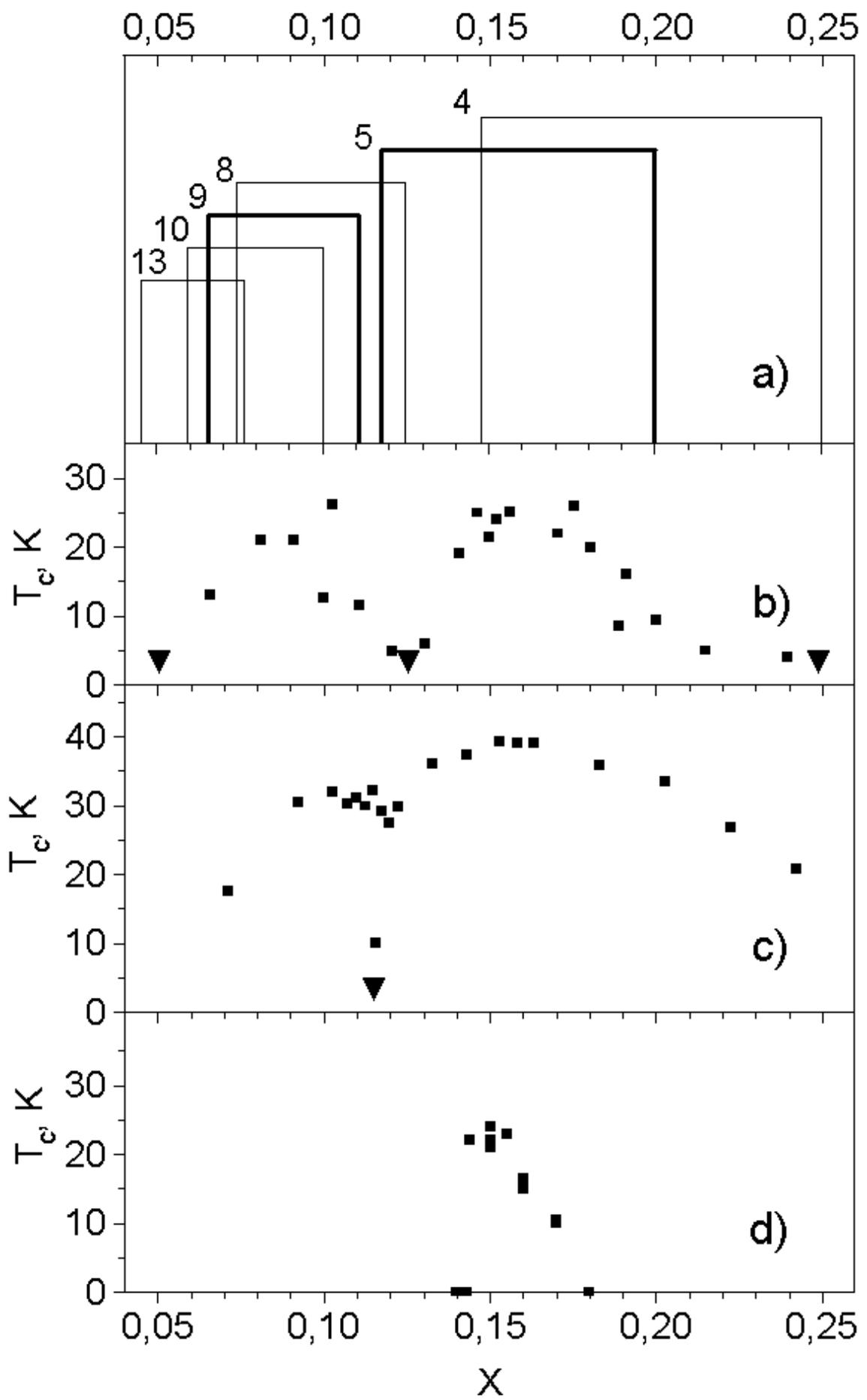
\end{document}